\documentclass[letterpaper, 10 pt, conference]{ieeeconf}  

\usepackage{graphicx}
\usepackage{tabularx}
\usepackage{multirow}
\usepackage{url}

\IEEEoverridecommandlockouts                              

\title{\LARGE \bf
Design of a GIS-based Assistant Software Agent  for the Incident Commander  to Coordinate  Emergency Response Operations 
}

\author{Reza Nourjou$^{1}$ \and Michinori Hatayama$^{1}$   \and  Stephen F. Smith$^{2}$ 
\and Atabak Sadeghi$^{3}$  \and Pedro Szekely$^{4}$
\thanks{$^{1}$Informatics Graduate School and DPRI, Kyoto University, Japan
        {\tt\small \{nourjour, hatayama\}@imdr.dpri.kyoto-u.ac.jp}}%
\thanks{$^{2}$The Robotics Institute, Carnegie Mellon University, USA
        {\tt\small sfs@cs.cmu.edu}}%
 \thanks{$^{3}$Geomatics Engineering Dep., Istanbul Technical University, Turkey     
      {\tt\small sadeghinaibin@itu.edu.tr}}%
 \thanks{$^{4}$Information Sciences Institute, University of Southern California, USA
        {\tt\small pszekely@isi.edu }}%
}

\begin{document}
\maketitle
\thispagestyle{empty}
\pagestyle{empty}

\begin{abstract}

Problem: This paper addresses  the design  of an intelligent software system for the IC (incident commander)  of a team in order to coordinate actions of agents-- field units or robots --in the domain of  emergency/crisis  response operations.

Objective: This paper proposes GICoordinator. It  is a GIS-based assistant software agent that assists  and collaborates with the human planner in strategic planning   and macro tasks assignment for centralized multi-agent coordination. 

Method: Our approach to design GICoordinator was to: analyze the problem, design a complete data model, design an architecture of GICoordinator, specify  required capabilities of human and system in coordination problem solving, specify development tools, and  deploy. 

Result: The result was an architecture/design  of GICoordinator that contains  system requirements.

Findings: GICoordinator efficiently integrates geo-informatics with artifice intelligent techniques  in order to provide a spatial intelligent coordinator system for an IC  to efficiently coordinate and control  agents  by making macro/strategic decisions. Results  define a framework for future works to develop  this system.

\end{abstract}

\section{INTRODUCTION} 

The domain of emergency/crisis response is concerned with reducing number of fatalities in the first few days after disaster (natural or human-made), and USAR (Urban Search and Rescue) has a significant issue in this domain.  A disaster response team, which contains an IC and several agents, is faced with the problem of carrying out geographically dispersed tasks under evolving execution circumstances in a manner that achieves a high-level objective in a minimum time. Agents need to efficiently coordinate their actions with each other in order to maximize the objective function. Effective coordination is an essential ingredient for  efficient emergency response management but it is difficult to achieve.

It is important for the IC, who has a big picture of the state of the world, to coordinate and control agents. His main role is to make a strategic action plan, allocate tasks to agents, decide on actions of agents, and schedule activities of agents in time and space.  

To propose an ideal system, some requirements should be considered as follows.  The  decision-maker is the IC, therefore coordination is a centralized approach but execution of tasks  is distributed among agents. Tasks information  that forms the IC`s global perception has the spatial, macro, dynamic, and temporal characteristic; it means that these data should be used in related algorithms.  Planning, task assignment, and  scheduling techniques in multi-agent systems are used for problem solving. Methods which are applied should partially constrain agents with macro decisions and permit agents to adapt their activities and make their own tactical (micro) decisions according to real situations. Because of the geographic characteristic of the problem,  GIS (geographic information
 systems) are required  to support human decisions by providing a set of  proper tools for management, analysis, modeling, and visualization of geographic information and location-based information. A mixed-initiative system can be proper system for the IC.

This requires an ideal intelligent software system to  assist the IC and collaborate with him in strategic planning and tasks assignment to agents according to the assessed requirements.

Although there is much literature \cite{vafaeinezhad2009using, Nourjou2011Introduction, maheswaran2011automated} in planning, coordination, task assignment, human-machine collaboration, problem-solving algorithms, and decision making under uncertainty,  unfortunately, the discussed  requirements  have not been thoroughly addressed by the previous works. 

Designing an ideal approach  is  an important phase in system development. As a result, in order to develop an ideal system, this paper aims to design  GICoordinator. GICoordinator is  a GIS-based assistant software agent that assists  and collaborates with the human planner in strategic planning   and macro tasks assignment in the mixed-initiative approach. This paper summarizes  phases that  make up this system.

\section{System Design}
This section is dedicated  to  phases which are important to design the GICoordinator.
 
\subsection{Analyze the Problem}
First step is to completely analyze and describe  the problem. Essential dimensions that compose this problem include: the problem domain, the structure of a team, geographic information,  macro tasks,  requirements, assumptions, the goal function, the strategic action plan, and the macro task schedule \cite{Nourjou2014Data}.

\subsection{Design the Architecture of  GICoordinator}
The architecture of GICoordinator is designed based on integration of three key components as Fig. 1 shows. Table 1 defines the required functionalities of GICoordinator and  human according to the analyzed problem. The following subsections briefly describe some important capabilities.

\begin{figure} [b!]
\centering
\includegraphics[width=.9\linewidth]{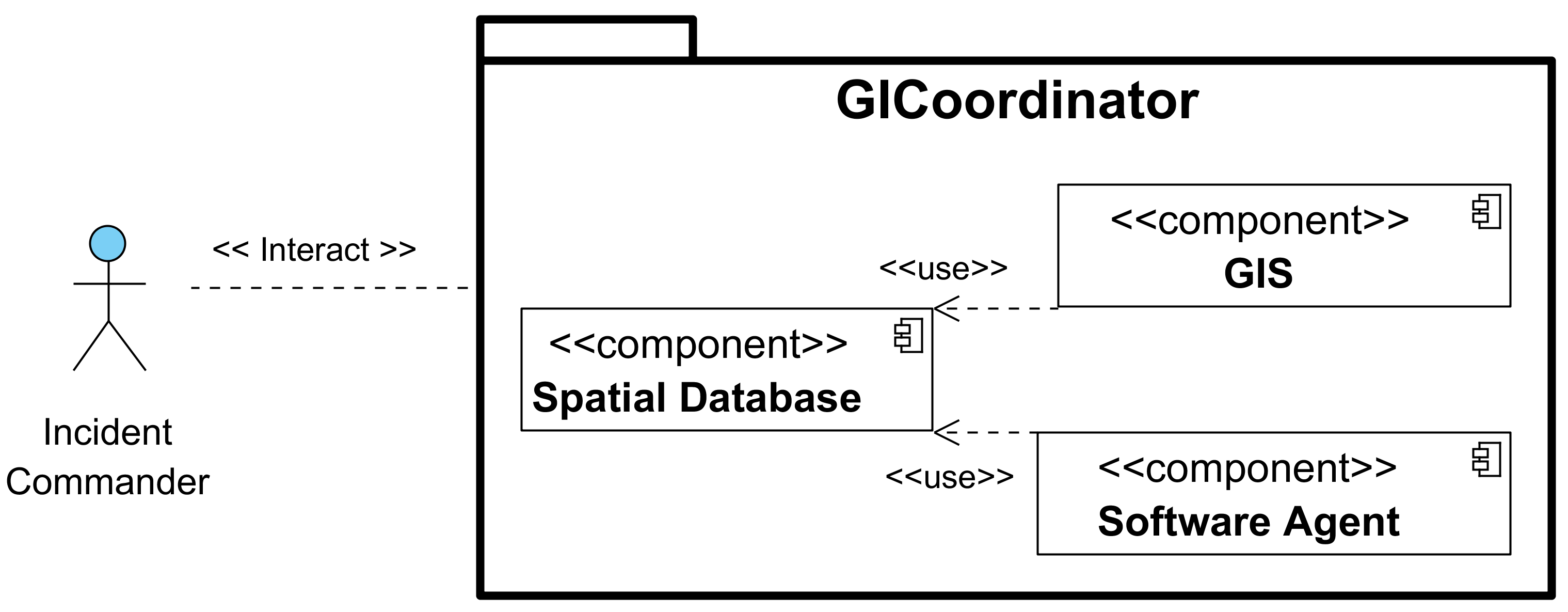}
\caption{The architecture of GICoordinator}
\label{Fig3}
\end{figure}	

\begin{table}[t]
\caption{System Requirements: the Desirable Capabilities of  GICoordinator and Human}
\label{}
\begin{center} 
\newcolumntype{c}[1]{>{\hsize=#1\hsize\raggedright\arraybackslash}X}%

\begin{tabularx}{.45\textwidth}{ |c{6.4}|c{19}|}  \hline
  \textbf{Component} & \textbf{Functionalities} \\ \hline
  
  \multirow{2}{*}{Human} & 1- Specify high-level strategies for coordination of agents \\ 
                                       &  2- Revise and refine strategies\\ 
                   \hline  
 \multirow{4}{*}{Spatial Database} &   3- Organize data (geographic, location-based, and non-geographic) of the problem\\ 
                                                      &   4- Integrate  the software agent with GIS\\ 
                                                      &   5- Support  information sharing with agents and with other information systems, and support  integration of information systems\\ 
               \hline    
 \multirow{5}{*}{GIS} & 6- Support human decisions by providing tools for geospatial reasoning, geographic information management and visualization \\ 
              			      & 7- Support the software agent by providing GIS analysis\\       
			                 & 8- Interact with human  \\                  
         										\hline      
\multirow{11}{*}{Software Agent} 
									& 9- Assign agents (field units) to human high-level strategies in strategic  planning  \\ 
  							        	& 10- Assign geospatial-temporal macro tasks to agents in centralized scheduling\\ 
  									& 11- Adapt and revise the strategic decision \\ 
									& 12- Adapt and revise the schedule \\ 
						                 & 13- Search for an optimal plan \\ 
						                 & 14- Allocate resources (refuges) to damage points under human strategies \\
						                 & 15- Adjust and refine human strategies\\
						                 & 16- Provide an interface to interact with human\\
                                              & 17- Percept and observe the environment via the spatial database\\
						                 						                     \hline        
 \end{tabularx}
\end{center}
\end{table}
 
 \subsubsection{Provide GIS Analysis}
 
 GICoordinator provides geo-technologies and geo-informatics that support human decisions and increase functionality of  the software agent. Three significant purposes are of importance: (1) visualization of  information via   3D maps or  thematic maps e.g.  display spatial distribution of tasks, states of macro tasks,  action plans,  and  tasks schedule, (2) geospatial reasoning that includes spatial relationships analysis, network analysis, proximity analysis, etc,   and (3) information management that includes retrieval, insert, query, update, and manipulate  information of spatial database. It enables the IC to interact with geographic information, and it supports human decisions by providing a better situational awareness. 

 GICoordinator requires necessary information for computation because it is mainly focus on multi-agent coordination techniques. Different types of information are provided by distributed systems. Therefore GICoordinator requires another capability to gather, integrate, mine, or fuse data from distributed databases. This version of GICoordinator does not include this functionality. 

 \subsubsection{Strategic Planning}

It is an approach in an organizational structure to make a  strategic plan that  states that how agents  can get from the current state of the world through a sequence of actions to a desired goal state. Strategic planning in a disaster response team includes (1) specify a response objective (a high-level strategy) for the team and decompose it into prioritized sub-goals (threads), (2) make a strategic decision by assignment of agents to threads, and (3) evaluate and adapt a strategic decision to new crisis situations \cite{maheswaran2011automated}.

This  capability  calculates a set of right choices for making a strategic decision either in execution of  the human strategy or in adaption of  current assignments \cite {Nourjou2014Intelligent}. These choices  are presented for the IC to select the best choice according to his intuition or to delegate the system to search for the optimal one. Moreover, the system  autonomously  releases right agents from a right thread in a right time during tasks execution in order to revise the strategic decision.

\subsubsection{Centralized Scheduling}
This capability is required to dynamically assign spatial-temporal  macro tasks to agents under  human strategic decisions in centralized scheduling in order to minimize the overall time of tasks execution. Two main results are achieved by running this  algorithm: (1) a feasible schedule and (2) an adaption time.  A schedule is composed of a number of macro decisions that specify: (1) what task type is going to be done, (2) who (a subset of agents) are assigned to do this assignment, (3) where (a  macro geographic object) contains a subset of tasks, (4) when operations start, (5) when operations finish, (6) how many tasks are estimated to be done, and (7) what task types and how many of them are estimated to be revealed in this location after to finish this job \cite{Nourjou2014Dynamic}.

\subsubsection{State-Space Search}
This capability calculates an optimal  strategic plan, a      complete  schedule, and a overall  minimum time of tasks execution. It evolves assignment of agents to threads from an initial state to a specified goal state. Results state  that  how and when agents  can reach a defined objective. These results support human decisions and assist the IC to evaluate the quality of his  strategy, refine it, or define a better strategy.

\subsubsection{Resource Allocation}
This capability optimally allocates  available refuges to rescued people in medical  transportation operations according to human strategic decisions. Results state that which injured persons should be transported to which refuges in order to optimize an objective function.

\subsubsection{Adjustment of Human Strategies}
It is difficult  for an IC to specify a good  strategy and timely revise it during emergency management. An IC may specify a bad or wrong high-level strategy that leads to a very big catastrophe with significantly severe consequences.  This capability  recommends human for adjustment and refinement of  his strategy in real-time. 

System should resolve trade-off of resource/role assignments   and reflect feed back from actual damaged area. This system requirement should consider them. Machine learning algorithms can provide proper solution. We will devote a paper to this functionality.

\subsection{Design the Data Model}
It is important to  model, formulate, and present data of the problem completely. The data model  presents elements of this problem,  properties, relationships, and interaction among these elements with regard to problem data modeling.  This  data model is important  to support development of  GICoordinator and implementation of the capabilities designed for this system \cite{Nourjou2014Data}.

 The designed data model only formulates the planning \& scheduling problem which the IC is faced, although spatially distributed agents can   observe, gather, and report many types of information  from their local environment to the incident center. Uncertainties are presented  by proper attributes  of  classes of this data model, and they are used by related algorithms.   
 
 This data model  formulates and present  any  USAR operations and similar crisis operations. It, also, formulates   a team  of different agents. The size of a team and the size of disaster scenario are  scalable, but it is important to model any required information in the data model. 

\subsection{Development tools}
There are many tools and approaches  that can be  used to implement and develop a system. In our methodology,  Table 2 shows required tools that  were used to develop GICoordinator. It states that how each component should be implemented and developed. 

We applied the  C\#.Net  programming language  for developing GICoordinator. Algorithms, the structure of the system,  and rules were implemented in the program. All system requirements, which are defined in Table 1, were implemented by the developer using the specified development tools.

There is a key concern  in development of GICoordinator. The designed data model was used to develop his system, and there are interdependencies among functions of the system. All functions should be run in an integrated system. Any change in the data model may cause that the system does not work. 

\begin{table}[t]
\caption{ Requirement Tools for Development of the GICoordinator}
\label{}
\begin{center} 
\newcolumntype{L}[1]{>{\hsize=#1\hsize\raggedright\arraybackslash}X}%

\begin{tabularx}{.445\textwidth}{ | L{9} | L{16} |  }
\hline
\textbf{Component} & \textbf{Tools} \\ \hline

Spatial Database & Microsoft Spatial SQL Server, GeoDatabase, the  data model \\ \hline
GIS & ArcGIS, .NET Programming  \\ \hline      
Software Agent & C\#.NET Programming, ArcObjects, the data model \\ \hline
         
\end{tabularx}
\end{center}
\end{table}

\subsection{Deploy}
GICoordinator is deployed in two ways.  First one is to calculate an feasible  strategic plan \& schedule, which partially specify actions of agents, by the human-system collaboration before execution. Second one is to use this system for automated and autonomous adaption/refinement of macro/strategic decisions to new situations during execution and in real time. 

As Fig. 1 shows, the GIS and the software agent have a connection to the spatial database. In oder to run  GICoordinator, the spatial database, whose structure is based on  the designed data model,  has to contain essential  information of the initial state of the world. These data can  insert or modified by the human via the user-interface that   GICoordinator provides for human-system interaction \cite{Nourjou2014Data}.

\section{Results}

The result was an architecture/design  of GICoordinator that contains  system requirements.   
  
\section{CONCLUSION}
The design  of GICoordinator was discussed in this paper. The key insight is  (1) support human decisions with geo-spatial intelligent software system, (2) provide  A.I. techniques for strategic planning, macro tasks assignment, scheduling, and automated adaption of these decisions in central multi-agent coordination, (3) involve human in the loop and  enable   collaboration between human and system for decision making. Future works will be to address the defined capabilities of GICoordinator by several papers.

\section*{ACKNOWLEDGMENT}
R.N. is grateful for the financial support of GCOE-HSE of Kyoto University, which enabled him to be a visiting scholar at the Information Sciences Institute of University of Southern California and the Robotics Institute of the Carnegie Mellon University during Dec. 2011 and Nov. 2012.

\end{document}